\documentclass[aps,floats]{revtex4}
\usepackage{amsmath,amssymb}
\usepackage{graphicx,epsfig}

\begin{document}
\bibliographystyle {plain}

\def\oppropto{\mathop{\propto}} 
\def\opsimeq{\mathop{\simeq}}
\def\opoverderline{\mathop{\overline}}
\def\operarrow{\mathop{\longrightarrow}}
\def\opsim{\mathop{\sim}}

\def\fig#1#2{\includegraphics[height=#1]{#2}}
\def\figx#1#2{\includegraphics[width=#1]{#2}}


\title{ Zero-temperature spinglass-ferromagnetic transition : \\ scaling analysis of the domain-wall energy
  } 


 \author{ C\'ecile Monthus and Thomas Garel }
  \affiliation{Institut de Physique Th\'{e}orique,\\
  CNRS and CEA Saclay \\
 91191 Gif-sur-Yvette, France}

\begin{abstract}
For the Ising model with Gaussian random coupling of average $J_0$ and unit variance, the zero-temperature spinglass-ferromagnetic transition as a function of the control parameter $J_0$ can be studied via the size-$L$ dependent renormalized coupling defined as the domain-wall energy $J^R(L) \equiv  E_{GS}^{(AF)}(L)-E_{GS}^{(F)}(L)$ (i.e. the difference between the ground state energies corresponding to AntiFerromagnetic and Ferromagnetic boundary conditions in one direction). We study numerically the critical exponents of this zero-temperature transition within the Migdal-Kadanoff approximation as a function of the dimension $d=2,3,4,5,6$. We then compare with the mean-field spherical model. Our main conclusion is that in low dimensions, the critical stiffness exponent $\theta^c$ is clearly bigger than the spin-glass stiffness exponent $\theta^{SG}$, but that they turn out to coincide in high enough dimension and in the mean-field spherical model. We also discuss the finite-size scaling properties of the averaged value and of the width of the distribution of the renormalized couplings.

\end{abstract}

\maketitle

\section{ Introduction  }

Among the various phase transitions that occur in disordered systems,
the case of {\it zero-temperature } critical points is especially interesting
since a new critical droplet exponent $\theta^c$
 is present with respect to thermal transitions and modifies the standard hyperscaling relation.
This phenomenon has been studied in detail for the random field Ising model 
(see the review \cite{nattermann} and references therein).
For the random bond Ising model
 \begin{eqnarray}
H =  - \sum_{<i,j>} J_{i j} S_i S_j 
\label{defsg}
\end{eqnarray}
where the random couplings $J_{ij} $ are drawn with a Gaussian distribution of average $J_0$ and variance unity
 \begin{eqnarray}
P(J_{i j}) = \frac{1}{\sqrt{2 \pi}} e^{- \frac{(J_{i j}-J_0)^2}{2} }
\label{gauss}
\end{eqnarray}
there also exists also a zero-temperature transition as a function of the control parameter $J_0$ (see for instance \cite{southern_young,grinstein,grassberger,bendisch,rieger,simkin,hartmann,krzakala,amoruso_hartmann,liers,picco,melchert_hartmann,toldin,ceccarelli,thomas_katzgraber} and references therein).
Of course, many other works on the random bond Ising model 
have been devoted to the {\it thermal } phase transitions between
the spin-glass phase and the paramagnetic phase, or between the
ferromagnetic phase and the paramagnetic phase, but these transitions at finite
temperature will not be discussed
in the following, where we focus on zero temperature.

Within the droplet scaling theory \cite{mcmillan,bray_moore,fisher_huse}, this 
zero-temperature transition between the spin-glass order and the ferromagnetic order
can be studied via the properties of the size-dependent effective renormalized coupling
$J^R(L)$. For a $d$-dimensional disordered sample of linear size $L$ containing $N=L^d$ spins,
it is defined by the following Domain-Wall Energy
\begin{eqnarray}
J^R(L) \equiv  E_{GS}^{(AF)}(L)-E_{GS}^{(F)}(L) 
\label{defjr}
\end{eqnarray}
where $E_{GS}^{(AF)}(N)$ and $E_{GS}^{(F)}(N)$ are the ground state energies
corresponding to AntiFerromagnetic and and Ferromagnetic boundary conditions in the first direction respectively (the other $(d-1)$ directions keep periodic boundary conditions).
The two phases can be characterized via the scaling of the average value $J^R_{av}(L)$
and of the width $\Delta J^R(L) $ of the probability distribution of the 
renormalized coupling $J^R(L)$ over the disordered samples :

(i) in the spin-glass phase $J_0<J_c$, the average value $J^R_{av}(L)$
becomes negligible with respect to the width $\Delta J^R(L) $
\begin{eqnarray}
\frac{J^R_{av}(L)}{\Delta J^R(L) } \operarrow_{L \to +\infty} 0
\label{ratiosg}
\end{eqnarray}

(ii) in the random ferromagnetic phase $J_0>J_c$, the width $\Delta J^R(L) $
becomes negligible with respect to the average value $J^R_{av}(L)$
\begin{eqnarray}
\frac{J^R_{av}(L)}{\Delta J^R(L) } \operarrow_{L \to +\infty} + \infty
\label{ratioferro}
\end{eqnarray}

(iii) at the critical point $J_0=J_c$, the averaged value  $J^R_{av}(L)$
and the width $\Delta J^R(L) $ remain in competition at all scales
\begin{eqnarray}
\frac{J^R_{av}(L)}{\Delta J^R(L) } \operarrow_{L \to +\infty} cst
\label{ratiocriti}
\end{eqnarray}

As emphasized by Bray and Moore on the example of the Random-Field Ising model \cite{BM_rfim},
the presence of some critical droplet exponent $\theta^c$ is directly related to the 
zero-temperature nature of the fixed point : for a thermal fixed point occurring at a finite $T_c$,
the fixed point corresponds to a fixed ratio $J_L/T_c$ so that the renormalized coupling $J_L$ has to be
 invariant under a change of scale $L$ at criticality ; for zero-temperature fixed points however, 
the competition is not between some renormalized coupling $J_L$ and the temperature $T$, 
but between two types of renormalized couplings, in our present case the average value $ J^R_{av}(L)$
and the width $\Delta J^R(L) $. At the critical point, even if the ratio of Eq. \ref{ratiocriti}
is fixed, both are actually expected to grow as $L^{\theta^c}$ with the scale $L$.
Since the critical droplet exponent is positive $\theta^c>0$, the temperature $T$ is
 irrelevant with respect to the growing renormalized couplings
 $ J^R_{av}(L) \sim \Delta J^R(L)  \sim L^{\theta_c}$.
So from the point of view of the renormalization flows, 
this zero-temperature critical point is repulsive in the direction of 
the parameter $J_0$ that controls the transition between the spin-glass phase and the ferromagnetic phase,
 but is attractive in the temperature direction. As a consequence, it is expected to govern also the critical behaviors between the spin-glass phase and the ferromagnetic phase in a finite temperature region around $T=0$.

The aim of this paper is to study the critical exponents governing
 the averaged value $J^R_{av}(L) $ and the width $\Delta J^R(L) $.
The paper is organized as follows. In Section \ref{sec_mk}, we study numerically this zero-temperature transition within the Migdal-Kadanoff approximation as a function of the dimension $d=2,3,4,5,6$. In section \ref{sec_sph}, we analyze the mean-field spherical spin-glass model. Our conclusions are summarized in section \ref{sec_conclusion}. In Appendix \ref{app_rem}, we also recall the properties of this zero-temperature transition for Derrida's Random Energy Model.

\section{ Migdal-Kadanoff renormalization in dimensions $2 \leq d \leq 6$  }

\label{sec_mk}

\subsection{Renormalization equation for the renormalized coupling }

\begin{figure}[htbp]
\includegraphics[height=6cm]{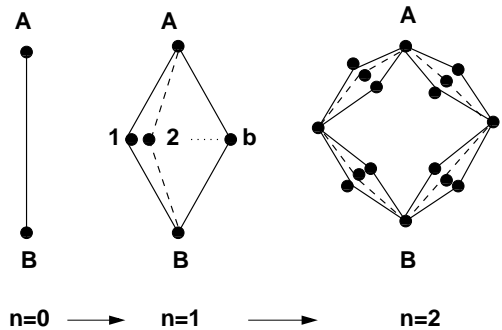}
\hspace{1cm}
\caption{ Hierarchical construction of the diamond lattice of
branching ratio $b$.   }
\label{figdiamond}
\end{figure}

Among real-space renormalization procedures \cite{realspaceRG}, 
Migdal-Kadanoff block renormalizations \cite{MKRG} play a special role
because they can be considered in two ways, 
 either as approximate renormalization procedures on hypercubic lattices,
or as exact renormalization procedures on certain hierarchical lattices
\cite{berker,hierarchical}.
One of the most studied hierarchical lattice is the
diamond lattice which is constructed recursively
from a single link called here generation $n=0$ (see Figure \ref{figdiamond}):
 generation $n=1$ consists of $b$ branches, each branch
 containing $2$ bonds in series ;
 generation $n=2$ is obtained by applying the same transformation
to each bond of the generation $n=1$.
At generation $n$, the length $L_n$ between the two extreme sites
$A$ and $B$ is $L_n=2^n$, and the total number of bonds is 
\begin{eqnarray}
B_n=(2b)^n = L_n^{d_{eff}(b)} \ \ \ {\rm \ \ with \  \ \  } 
d_{eff}(b)= \frac{ \ln (2b)}{\ln 2}
\label{deff}
\end{eqnarray}
where $d_{eff}(b)$ represents the fractal dimensionality.
On this diamond lattice, various disordered spins models have been studied,
such as the diluted Ising model \cite{diluted}, 
ferromagnetic random Potts model \cite{Kin_Dom,Der_Potts,andelman,us_potts,turban,igloi} and spin-glasses \cite{southern_young,young,mckay,Gardnersg,bray_moore,banavar,bokil,muriel,thill,boettcher,nishimori,jorg}.

Here we are only interested into the zero-temperature ground state energies given
the two boundary spins, which evolve according to the recursion
 \begin{eqnarray}
E_{GS}^{S_A,S_B} = \sum_{i=1}^b {\rm min} \left[ E_{GS}^{S_A,S_i=+1}+E_{GS}^{S_i=+1,S_B} ;
E_{GS}^{S_A,S_i=-1}+E_{GS}^{S_i=-1,S_B} \right]
\label{rgegsab}
\end{eqnarray}
with the initial condition
 \begin{eqnarray}
E_{GS}^{S_A,S_B} =-J_{AB} S_A S_B
\label{iniegs}
\end{eqnarray}
To take into account the invariance by a global flip of all the spins,
it is more convenient to introduce the ground state energies corresponding to
Ferromagnetic and AntiFerromagnetic boundary conditions
 \begin{eqnarray}
E_{GS}^{F} && \equiv E_{GS}^{+,+} =E_{GS}^{-,-} \nonumber \\
E_{GS}^{AF} && \equiv E_{GS}^{+,-} =E_{GS}^{-,+}
\label{defefeaf}
\end{eqnarray}
with the renormalization rules
 \begin{eqnarray}
E_{GS}^{F} && = \sum_{i=1}^b {\rm min} \left[E_{GS}^{F}(i_A)+E_{GS}^{F}(i_B)  ;
E_{GS}^{AF}(i_A)+E_{GS}^{AF}(i_B)\right]\nonumber \\
E_{GS}^{AF} && = \sum_{i=1}^b {\rm min} \left[E_{GS}^{F}(i_A)+E_{GS}^{AF}(i_B)  ;
E_{GS}^{AF}(i_A)+E_{GS}^{F}(i_B)\right]
\label{rgegsfaf}
\end{eqnarray}
where the notations $i_A$ and $i_B$ refers to the bonds of the branch $i$ connected respectively to the boundary $A$ and $B$ on the central lattice of Figure \ref{figdiamond}.

The corresponding initial conditions (Eq. \ref{iniegs}) read
 \begin{eqnarray}
E_{GS}^{F} =-J_{AB} \nonumber \\
E_{GS}^{AF} =J_{AB}
\label{iniegsfaf}
\end{eqnarray}
So the renormalized coupling of Eq. \ref{defjr}
 \begin{eqnarray}
J^R \equiv \frac{E_{GS}^{AF}-E_{GS}^{F}}{2}
\label{defJR}
\end{eqnarray}
evolves according to the well known renormalization rule
 \begin{eqnarray}
J^R = \sum_{i=1}^b {\rm sign} ( J_{i_A} J_{i_B})  {\rm min} \left[ \vert J_{i_A} \vert ;\vert J_{i_B} \vert \right]
\label{rgJR}
\end{eqnarray}

\subsection{ Numerical pool method }

\begin{figure}[htbp]
\includegraphics[height=6cm]{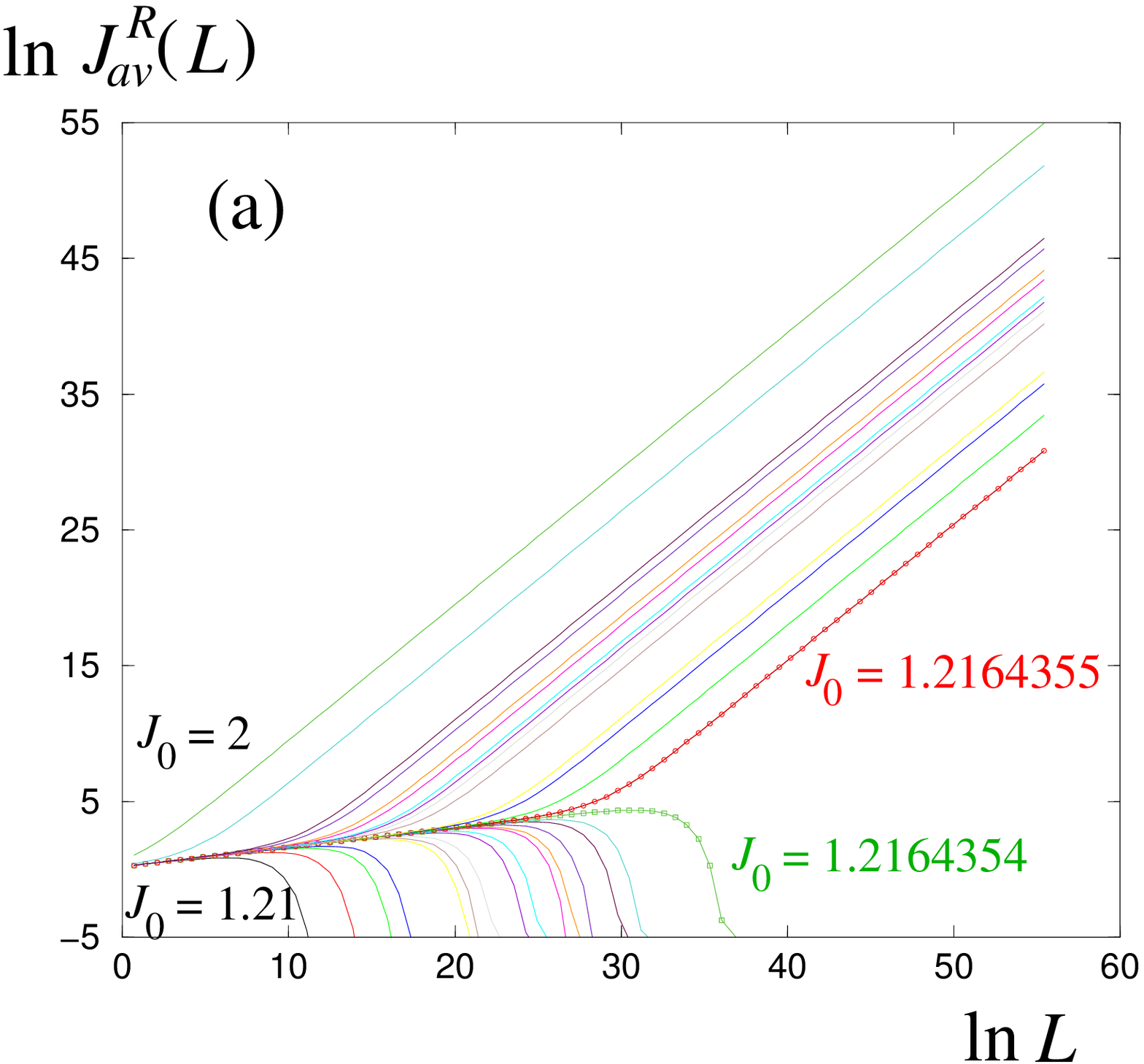}
\hspace{2cm}
\includegraphics[height=6cm]{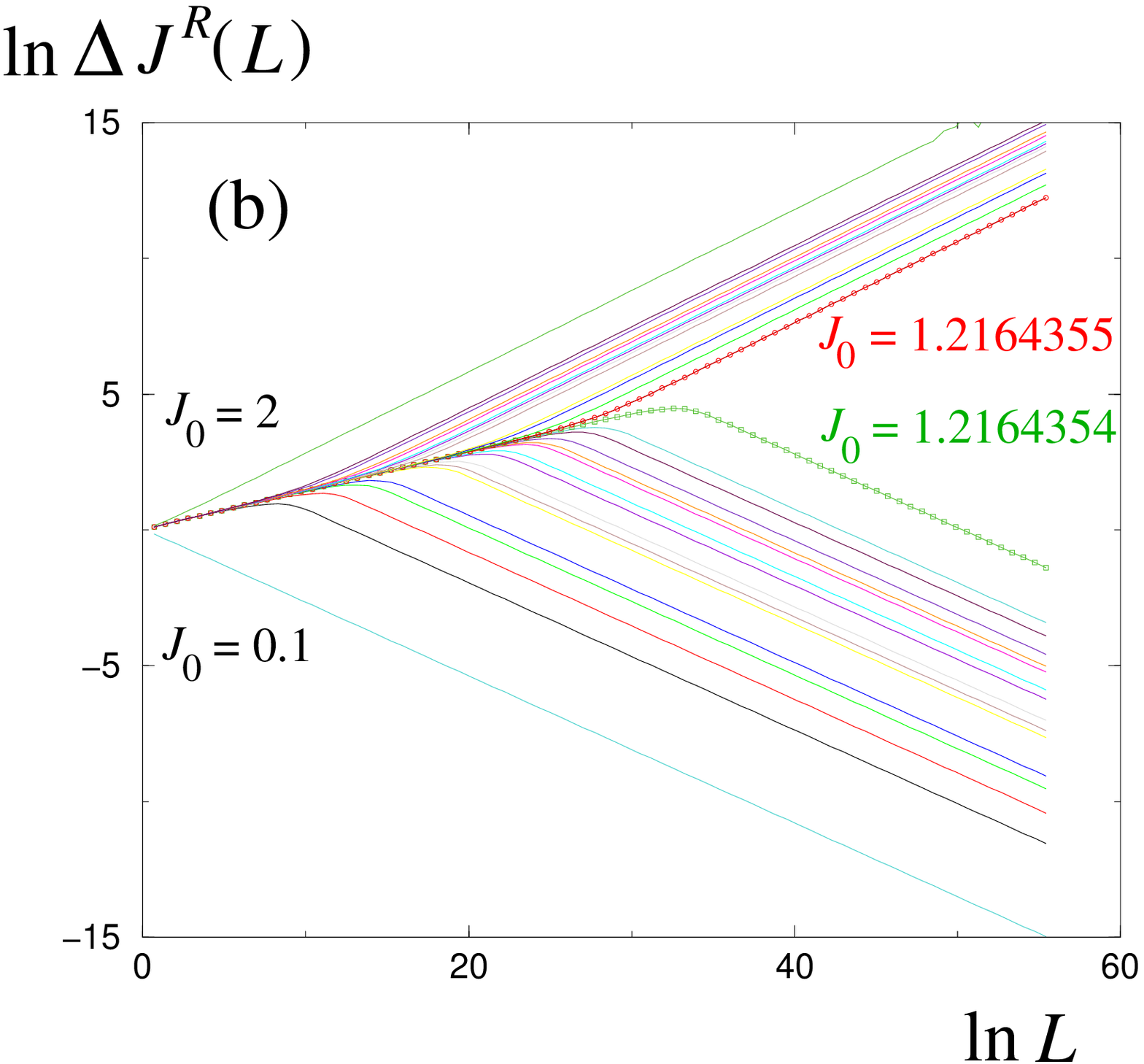}
\vspace{1cm}
\caption{ RG flows in log-log for the case $d_{eff}=2$ \\
(a) RG flow of the averaged value $J_{av}^R(L)$ as a function of the length $L$
for various initial values $J_0$ : \\
for $J_0 \geq (J^c_{pool})^+=1.2164555$, the asymptotic straight lines correspond to Eq. \ref{avJferro} :
$\ln J^R_{av} \simeq d_s \ln L+\ln \sigma(J_0)$, \\
whereas for $J_0 \leq (J^c_{pool})^-=1.2164554$, the averaged value $J^R_{av} $ flows towards zero. \\
(b) RG flow of the width $\Delta J^R (L)$ as a function of the length $L$
for various initial values $J_0$ : \\
for $J_0 \geq (J^c_{pool})^+=1.2164555$, the asymptotic straight lines correspond to Eq. \ref{wJferro} :
$\ln \Delta J^R \simeq \theta^{F}_{var} \ln L + \ln \rho(J_0)$, \\
whereas for $J_0 \leq (J^c_{pool})^-=1.2164554$, the asymptotic straight lines correspond to Eq. \ref{wJsg} :
$ \ln \Delta J^R \simeq \theta^{SG}\ln L+\ln \Upsilon(J_0)$.
}
\label{figrgflow}
\end{figure}

The standard method to study numerically the renormalization Eq \ref{rgJR}
is the so-called 'pool method'. 
The idea is to represent the probability distribution
$P_n(J_n)$ of the renormalized coupling $J_n$ at generation $n$
 by a pool of $M$ realizations $(J_n^{(1)},J_n^{(2)},...,J_n^{(M)})$.
The pool at generation $(n+1)$ is then obtained as follows :
each new realization $(J_{n+1}^{(i)})$ is obtained by choosing 
$(2 b)$ realizations at random from the pool of generation $n$ and by applying
the renormalization equations given in Eq. \ref{rgJR}.
The initial condition at generation $n=0$ corresponds to the Gaussian distribution of Eq. \ref{gauss}.
The numerical results presented below have been obtained with a pool of size
$M=3.10^7$ which is iterated up to $n=60$ or $n=80$ generations.
For each value of the fractal dimension $d_{eff}$ of Eq. \ref{deff},
this pool method is applied for various values of the control parameter $J_0$
of the initial condition in order to locate by dichotomy
the critical value $J_c$.

\subsection{ Critical point $J_0=J_c$ }

As an example on Fig. \ref{figrgflow} concerning the case $d_{eff}=2$,
we show our data concerning the RG flows of the
averaged value $J_{av}^R(L)$ and of the width $\Delta J^R (L)$ as a function of
the length $L$ for various initial values $J_0$ :
the pool critical parameter $J^c_{pool}$
is determined as the point where we see the bifurcations in both flows.
Our numerical data yield the following values 
 \begin{eqnarray}
J^{c}_{pool}(d_{eff}=2) && \simeq 1.21643545 \nonumber \\
J^{c}_{pool}(d_{eff}=3) && \simeq 0.66448034 \nonumber \\
J^{c}_{pool}(d_{eff}=4) && \simeq 0.419828938 \nonumber \\
J^{c}_{pool}(d_{eff}=5) && \simeq 0.281555071 \nonumber \\
J^{c}_{pool}(d_{eff}=6) && \simeq 0.193923375
\label{jcdiamond}
\end{eqnarray}

\begin{figure}[htbp]
\includegraphics[height=6cm]{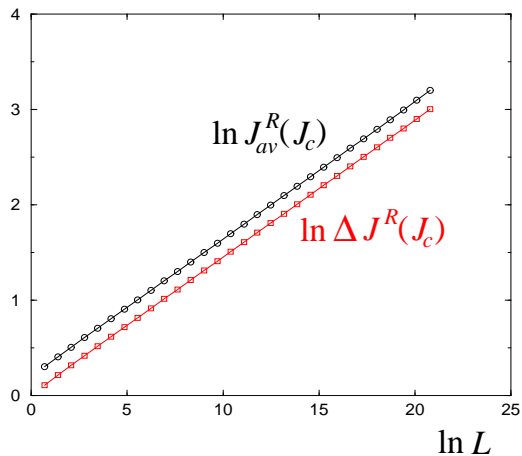}
\vspace{1cm}
\caption{ For $d_{eff}=2$,  comparison of the RG flows of the
averaged value $J_{av}^R(L)$ and of the width $\Delta J^R (L)$  at the critical point $ J^c_{pool}$
 [These critical RG flows are obtained by considering the RG flows for the values
$J_0=(J^c_{pool})^+=1.2164555$ and $(J^c_{pool})^-=1.2164554 $ as long as they coincide
before the bifurcations shown on Fig. \ref{figrgflow}].
 The common slope of the two RG flows yields the critical droplet exponent
$\theta_c $ of Eq \ref{jcriri}. }
\label{figcritiflow}
\end{figure}

For length scales before the bifurcation, the RG flows concerning the nearest
values of $J^c_{pool}$ from above $(J^c_{pool})^+$ and from below $(J^c_{pool})^-$ coincide
and represent the critical RG flows. As expected, and as 
as shown on Fig. \ref{figcritiflow} for the case $d_{eff}=2$, 
the critical flows of the averaged value  $J^R_{av}(L)$
and the width $\Delta J^R(L) $ are governed by the same exponent  $\theta^{c}$
(so that they remain in competition at all scales (Eq \ref{ratiocriti}))
 \begin{eqnarray}
 J^R_{av} (L;J_0=J_c) && \propto  L^{\theta^{c}}
\nonumber \\
\Delta J^R(L;J_0=J_c) && \propto L^{\theta^{c}}
\label{jcriri}
\end{eqnarray}

Within the Migdal-Kadanoff renormalization, our numerical results yields
the following value as a function of the dimension $d_{eff}$  of Eq. \ref{deff}
 \begin{eqnarray}
\theta^{c}(d_{eff}=2) && \simeq 0.14 \nonumber \\
\theta^{c}(d_{eff}=3) && \simeq 0.46 \nonumber \\
\theta^{c}(d_{eff}=4) && \simeq 0.87 \nonumber \\
\theta^{c}(d_{eff}=5) && \simeq 1.32  \nonumber \\
\theta^{c}(d_{eff}=6) && \simeq 1.79
\label{thetacdiamond}
\end{eqnarray}
Note that the value obtained here $\theta^{c}(d_{eff}=2) \simeq 0.143$
is actually close to the values measured on the square lattice, namely
$\theta^{c}(d=2)=0.19(2) $  and $\theta^{c}(d=2)=0.16(4) $ in Ref. \cite{rieger}
as well as $\theta^{c}(d=2)=0.12(5) $  and $\theta^{c}(d=2)=0.13(5) $ in Ref. \cite{amoruso_hartmann}. Unfortunately for hypercubic lattices in dimension $d=3,4,5,6$, we are not aware of numerical measures of the exponent $\theta^c$ to compare with the Migdal-Kadanoff values of Eq. \ref{thetacdiamond}.

It is also interesting to note that the critical exponents $\theta^c(d_{eff})$
of Eq. \ref{thetacdiamond} 
seem to be very close to the critical exponents
 $\theta_Z(d_{eff})$ given in Table 1 of Ref. \cite{igloi} concerning
 the disordered Potts model in the large-q limit on the same diamond lattices
where there is no spin-glass phase. It is not clear to us whether this is a coincidence or not.

\subsection{ Spin-Glass phase $J_0<J_c$ }

In the Spin-Glass phase $J_0<J_c$, 
the average value $J^R_{av}(L)$
becomes negligible with respect to the width $\Delta J^R(L) $ (Eq. \ref{ratiosg}).
More precisely, the averaged coupling vanishes asymptotically
(see Fig. \ref{figrgflow} (a))
 \begin{eqnarray}
 J^R_{av} (0 \leq J_0<J_c)  \operarrow_{L \to +\infty} 0
\label{avJsg}
\end{eqnarray}
whereas the width of the distribution grows with the stiffness exponent $\theta^{SG}$
(which coincides with the droplet exponent within the droplet scaling theory \cite{mcmillan,bray_moore,fisher_huse})
 \begin{eqnarray}
\Delta J^R (0 \leq J_0<J_c) \propto \Upsilon(J_0) L^{\theta^{SG}}
\label{wJsg}
\end{eqnarray}
as shown on Fig. \ref{figrgflow} (b) for the case $d_{eff}=2$.
Within the Migdal-Kadanoff approximation, our numerical results of the pool method for the diamond lattice yield the following values as a function of the fractal dimension
\begin{eqnarray}
\theta^{SG}(d_{eff}=2) && \simeq -0.27 \nonumber \\
\theta^{SG}(d_{eff}=3) && \simeq 0.26 \nonumber \\
\theta^{SG}(d_{eff}=4) && \simeq 0.76 \nonumber \\
\theta^{SG}(d_{eff}=5) && \simeq 1.27 \nonumber \\
 \theta^{SG}(d_{eff}=6) && \simeq 1.77
\label{thetasgdiamond}
\end{eqnarray}
that may be compared with the stiffness exponents measured on hypercubic lattices
(see \cite{boettcher_hyper} and references therein) : 
$\theta^{SG}(d=2)  \simeq - 0.28$; 
$\theta^{SG}(d=3)  \simeq 0.24$;
$\theta^{SG}(d=4)  \simeq 0.61$;
$\theta^{SG}(d=5)  \simeq 0.88$;
$\theta^{SG}(d=6) \simeq 1.1$.

\begin{figure}[htbp]
\includegraphics[height=6cm]{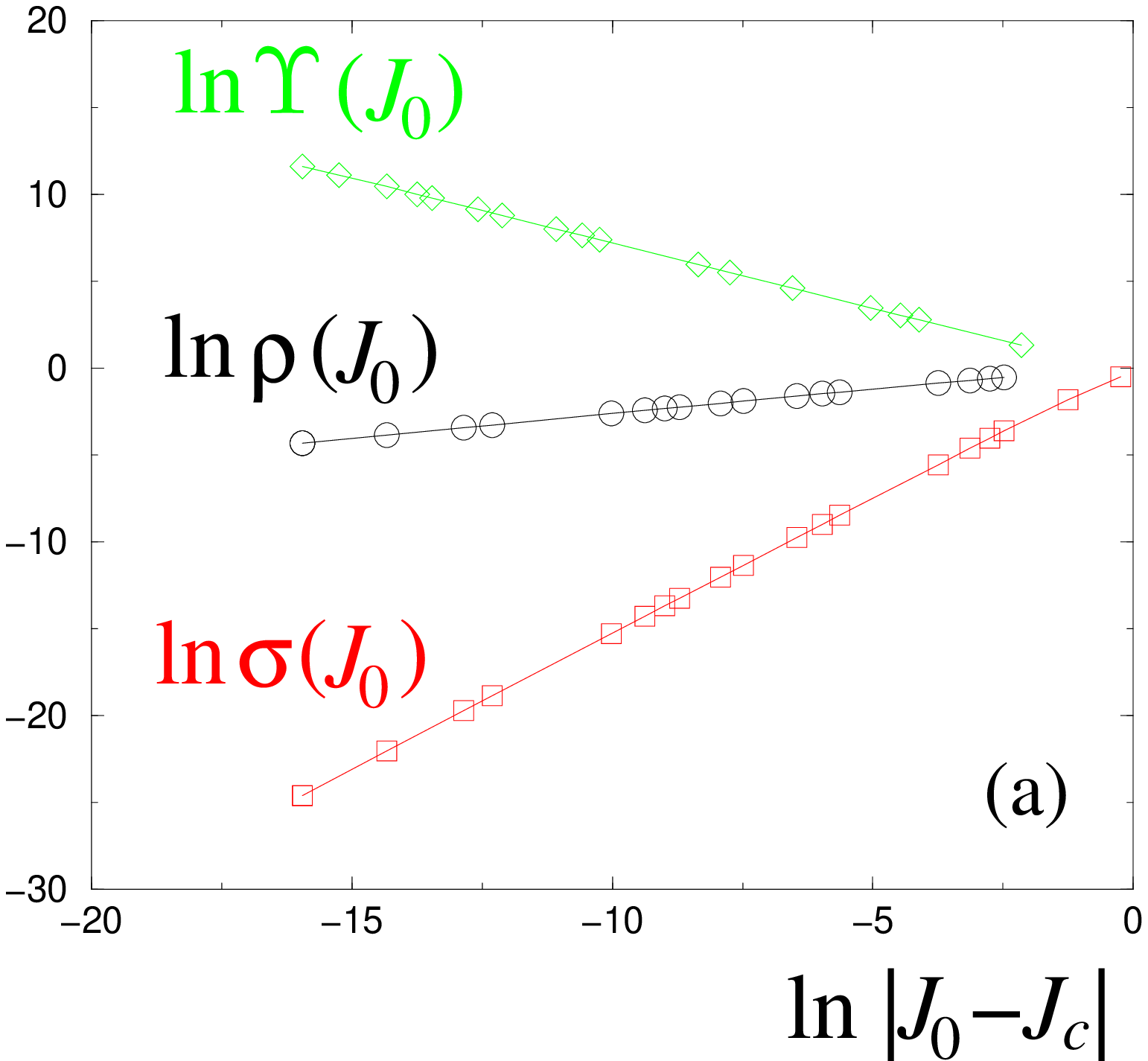}
\hspace{2cm}
\includegraphics[height=6cm]{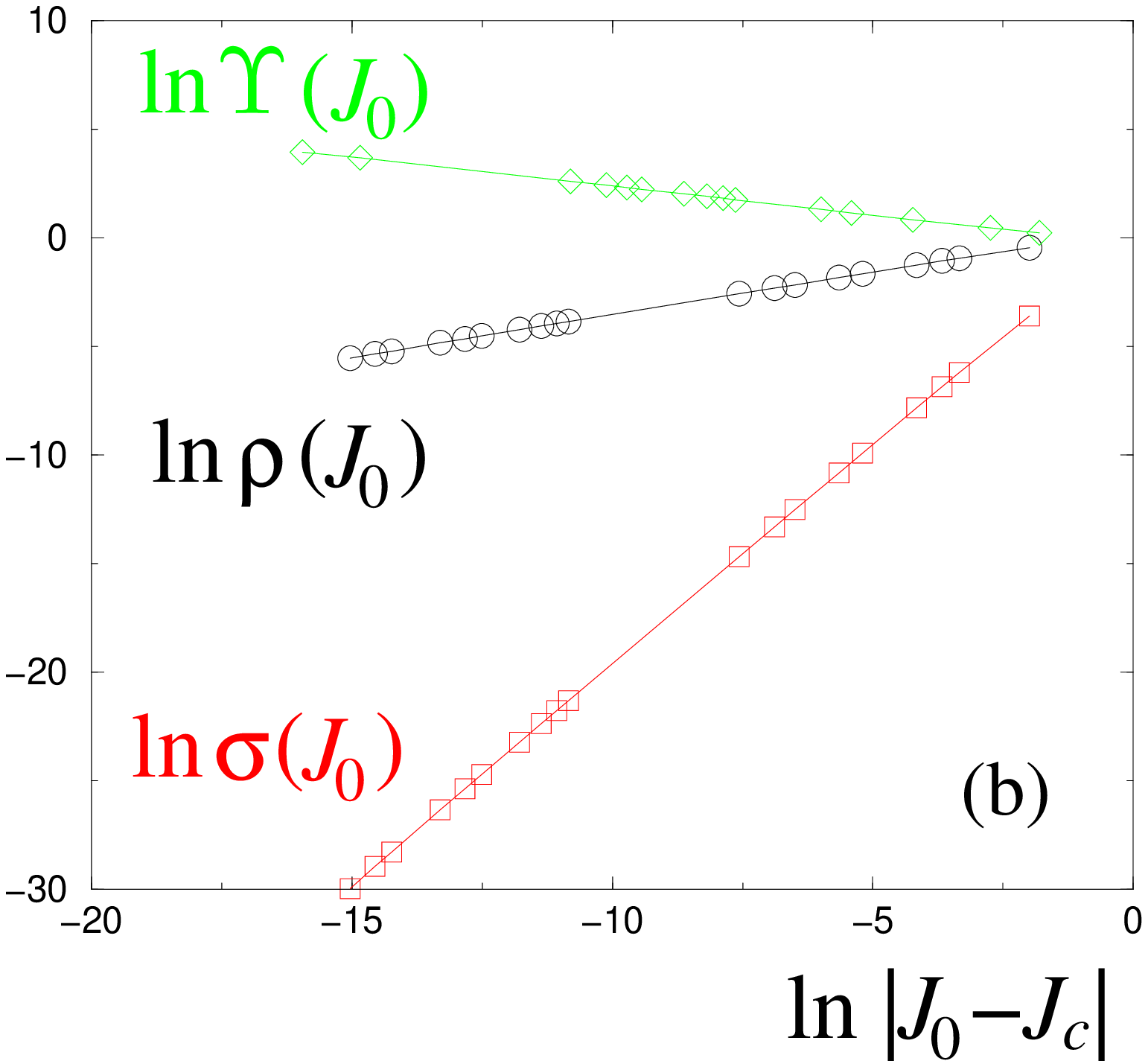}
\vspace{1cm}
\caption{Measure of the critical exponents concerning the stiffness modulus
$\Upsilon(J_0) \propto (J_c-J_0)^{-y}$ of the spin-glass phase $J_0<J_c$,
 the surface tension $\sigma(J_0) \propto (J_0-J_c)^{s} $ and the
amplitude $\rho(J_0) \propto (J_0-J_c)^{r}$
of the random ferromagnetic phase $J_0>J_c$ : \\
(a) for the case $d_{eff}=2$, the slopes yield $y \simeq 0.75$ ,$s \simeq
1.55$ and $r \simeq 0.28$ \\
(b) for the case $d_{eff}=3$, the slopes yield $y \simeq 0.27 $ , $s \simeq
2.02$ and $r \simeq 0.39$. \\
Results for other dimensions $d_{eff}$ are given in Eqs
\ref{dvupssgdiamond}, \ref{critisigmadiamond} and \ref{critirhodiamond} . }
\label{figexponents}
\end{figure}

From the asymptotic straight lines of the RG flow of the width $\Delta J^R $ for $J_0<J^c_{pool}$
(see Fig. \ref{figrgflow} (b))
we may extract the stiffness modulus $\Upsilon(J_0)$ and plot it as shown on Fig.
\ref{figexponents} to estimate the critical exponent $y$ governing the divergence
near the transition
 \begin{eqnarray}
\Upsilon(J_0<J_c) \oppropto_{J_0 \to J_c^-} (J_c-J_0)^{-y}
\label{dvupsilon}
\end{eqnarray}
Our numerical data yield the following values for the critical exponent $y$ 
 as a function of the fractal dimension
 \begin{eqnarray}
y(d_{eff}=2) && \simeq 0.75  \nonumber \\
y(d_{eff}=3) && \simeq 0.27  \nonumber \\
y(d_{eff}=4) && \simeq 0.12  \nonumber \\
y(d_{eff}=5) && \simeq 0.05  \nonumber \\
y(d_{eff}=6) && \simeq 0.02
\label{dvupssgdiamond}
\end{eqnarray}

Let us now consider the following finite-size scaling form \cite{rieger}
 \begin{eqnarray}
\Delta J^R(J_0<J_c) && \propto L^{\theta^{c}} \Phi_{SG} \left( \frac{L}{\xi^{SG}_{var}(J_0) }  \right)
\label{fsssgwidth}
\end{eqnarray}
to define the spin-glass correlation length 
 \begin{eqnarray}
\xi^{SG}_{var}(J_0)  \oppropto_{J_0 \to J_c^-} (J_c-J_0)^{-\nu^{SG}_{var}}
\label{nusg}
\end{eqnarray}
The matching with Eq \ref{wJsg} yields the following
power-law behavior $\Phi_{SG}(x) \propto x^{\theta^{SG}-\theta^c}$ at large $x$
yielding the following divergence for the stiffness modulus
 \begin{eqnarray}
\Upsilon(J_0<J_c) && \oppropto_{J_0 \to J_c^-} \left[\xi^{SG}_{var}(J_0)\right]^{\theta^c-\theta^{SG}}
\label{fsssgwidthasymp}
\end{eqnarray}
or equivalently the following relation between critical exponents
 \begin{eqnarray}
y = \nu^{SG}_{var} (\theta^c-\theta^{SG} )
\label{nusgvar}
\end{eqnarray}

The previous numerical results given for $\theta^c $, $\theta^{SG}$ and $y$ yield
 \begin{eqnarray}
\nu^{SG}_{var}(d_{eff}=2) && \simeq \frac{0.75}{0.14- (-0.27)} \simeq 1.8 \nonumber \\
\nu^{SG}_{var}(d_{eff}=3) && \simeq \frac{0.27}{0.46    - 0.26} \simeq 1.35
\nonumber \\
\nu^{SG}_{var}(d_{eff}=4) && \simeq   \frac{0.12 }{0.87-0.76} \simeq 1.1
\label{nusgvardiamond}
\end{eqnarray}
The two other cases $d_{eff}=5$ and $d_{eff}=6$ do not give precise estimations
of $\nu^{SG}_{var} $,
because the numerator and the denominator are both small.

\subsection{  Random Ferromagnetic phase $J_0>J_c$ }

 In the random ferromagnetic phase $J_0>J_c$, the width $\Delta J^R(L) $
becomes negligible with respect to the average value $J^R_{av}(L)$ (Eq. \ref{ratioferro}), 
but it is interesting to consider the behavior of both.

\subsubsection{ Averaged renormalized coupling } 

For $J_0>J_c$ (see Fig. \ref{figrgflow} (a)), the averaged renormalized coupling $J^R_{av} (J_0>J_c) $ presents the same scaling as the pure ferromagnet : for short-ranged models, the energy cost of an interface grows as the surface $L^{d_s}$ of a system-size interface
 \begin{eqnarray}
 J^R_{av} (J_0>J_c) \propto \sigma(J_0) L^{d_s}
\label{avJferro}
\end{eqnarray}
where the interface dimension is simply
 \begin{eqnarray}
 d_s=d_{eff}-1
\label{defdsSR}
\end{eqnarray}

Since the surface dimension is always bigger than the critical stiffness exponent
$\theta^c$ (Eq. \ref{thetacdiamond}), the surface tension $ \sigma(J_0)$
of Eq. \ref{avJferro} vanishes at the transition
 \begin{eqnarray}
\sigma(J_0>J_c) \oppropto_{J_0 \to J_c^+} (J_0-J_c)^{s}
\label{critisigma}
\end{eqnarray}

From the asymptotic straight lines of the RG flow for $J_0>J^c_{pool}$
(see Fig. \ref{figrgflow} (a) ),
we may extract the surface tension $\sigma(J_0)$ and plot it as shown on Fig.
\ref{figexponents} to estimate the critical exponent $s$ of Eq. \ref{critisigma}
 \begin{eqnarray}
s(d_{eff}=2) && \simeq 1.55  \nonumber \\
s(d_{eff}=3) && \simeq 2.02  \nonumber \\
s(d_{eff}=4) && \simeq 2.4 \nonumber \\
s(d_{eff}=5) && \simeq 2.85  \nonumber \\
s(d_{eff}=6) && \simeq 3.28
\label{critisigmadiamond}
\end{eqnarray}

The finite-size scaling form \cite{rieger}
 \begin{eqnarray}
 J^R_{av} (J_0>J_c) && \propto  L^{\theta^{c}} \Psi_F \left( \frac{L}{\xi^{F}_{av}(J_0) }  \right)
\label{fssferroav}
\end{eqnarray}
allows to define the ferromagnetic correlation length 
 \begin{eqnarray}
\xi^{F}_{av}(J_0)  \oppropto_{J_0 \to J_c^+} (J_0-J_c)^{-\nu^{F}_{av}}
\label{nuavf}
\end{eqnarray}
The matching with Eq.  \ref{avJferro}
yields the following behavior for the surface tension
 \begin{eqnarray}
 \sigma(J_0>J_c) && \propto   \left[\xi^{F}_{av}(J_0)\right]^{\theta^c-\theta^F_{av}}
\label{fssferroavasymp}
\end{eqnarray}
i.e. the following relation between critical exponents
 \begin{eqnarray}
s= \nu^{F}_{av}(d_s-\theta^c)
\label{nufav}
\end{eqnarray}
Note the difference with the Widom relation $s=\nu d_s=\nu (d-1)$ \cite{widom}
for thermal transition characterized by $\theta_c=0$.
The previous values given for $\theta_c$ and $s$ yield
 \begin{eqnarray}
\nu^{F}_{av}(d_{eff}=2) && \simeq \frac{1.55}{1-0.14} \simeq 1.8  \nonumber \\
\nu^{F}_{av}(d_{eff}=3) && \simeq \frac{2.02}{2-0.46  } \simeq 1.3
\nonumber \\
\nu^{F}_{av}(d_{eff}=4) && \simeq  \frac{2.4}{3-0.87} \simeq 1.13
\nonumber \\
\nu^{F}_{av}(d_{eff}=5) && \simeq  \frac{2.85 }{4-1.32} \simeq 1.06
\nonumber \\
\nu^{F}_{av}(d_{eff}=6) && \simeq  \frac{3.28}{5-1.79} \simeq 1.02
\label{nuFavdiamond}
\end{eqnarray}
that coincide from the estimate of Eq. \ref{nusgvardiamond} within the spin-glass phase.

\subsubsection{ Width of the probability distribution of the renormalized coupling  } 

Let us now consider the width $\Delta J^R (J_0>J_c)$ of the probability distribution of the renormalized coupling around the averaged value of Eq. \ref{avJferro} (see Fig. \ref{figrgflow} (b)) : it grows
with some exponent $\theta^{F}_{var}$ representing the droplet exponent of a directed
interface of dimension $d_s=d-1$ in a space of dimension $d$
 \begin{eqnarray}
\Delta J^R (J_0>J_c) \propto \rho(J_0) L^{\theta^{F}_{var}}
\label{wJferro}
\end{eqnarray}
This exponent $\theta^{F}_{var} $ takes the following values within the 
 Migdal-Kadanoff approximation
 \begin{eqnarray}
\theta^{F}_{var}(d_{eff}=2) && \simeq 0.3 \nonumber \\
\theta^{F}_{var}(d_{eff}=3) && \simeq 0.76 \nonumber \\
\theta^{F}_{var}(d_{eff}=4) && \simeq 1.24 \nonumber \\
\theta^{F}_{var}(d_{eff}=5) && \simeq 1.73  \nonumber \\
\theta^{F}_{var}(d_{eff}=6) && \simeq 2.23
\label{thetafdiamond}
\end{eqnarray}
Note that the value $ \theta^{F}_{var}(d_{eff}=2) \simeq 0.3$ corresponds to the
droplet exponent of the Directed Polymer model on the diamond lattice with $d_{eff}=2$
\cite{Der_Gri,us_DPdiamond,us_potts} that may be compared with the exact 
value $\theta^{F}_{var}(d=2) =\frac{1}{3}  $ for the Directed Polymer on the square lattice \cite{henley}. The values $\theta^{F}_{var}(d_{eff}=3) \simeq 0.76 $ and $\theta^{F}_{var}(d_{eff}=4)  \simeq 1.24 $ may be compared with the corresponding droplet exponents $\theta^{F}_{var}(d=3) \simeq 0.84 $ and $\theta^{F}_{var}(d=4)  \simeq 1.45 $ measured on hypercubic lattices \cite{middleton}.

It is also interesting to note that
 the critical exponents $\theta^F_{var}(d_{eff})$ 
of Eq. \ref{thetafdiamond}
 coincide up to numerical errors with the critical exponents $\theta(d_{eff})$
of the disordered Potts model in the large-q limit on the same diamond lattices
(see Table 1 of Ref. \cite{igloi}). More generally, the exponents $\theta^F_{var}(d_{eff})$ are expected to be the same for all values of the parameter $q$ of the Potts model (see also \cite{us_potts} where $ \theta^{F}_{var}(d_{eff}=2)  \simeq 0.3$ has also been measured for the Potts model of parameter $q=8$).

Since the fluctuation exponent $\theta^{F}_{var}  $ in the ferromagnetic phase
 is bigger than the critical stiffness exponent
$\theta^c$ (Eq. \ref{thetacdiamond}), the amplitude $ \rho(J_0)$
of Eq. \ref{wJferro} vanishes at the transition
 \begin{eqnarray}
\rho(J_0>J_c) \oppropto_{J_0 \to J_c^+} (J_0-J_c)^{r}
\label{critirho}
\end{eqnarray}
From the asymptotic straight lines of the RG flow for $J_0>J^c_{pool}$
(see Fig. \ref{figrgflow} (b) )
we may extract the amplitude $\rho(J_0)$ and plot it as shown on Fig.
\ref{figexponents} to estimate the critical exponent $r$
 \begin{eqnarray}
r(d_{eff}=2) && \simeq 0.28  \nonumber \\
r(d_{eff}=3) && \simeq 0.39  \nonumber \\
r(d_{eff}=4) && \simeq 0.43  \nonumber \\
r(d_{eff}=5) && \simeq 0.44   \nonumber \\
r(d_{eff}=6) && \simeq 0.45
\label{critirhodiamond}
\end{eqnarray}

The finite-size scaling form \cite{rieger}
 \begin{eqnarray}
\Delta J^R(J_0>J_c) && \propto L^{\theta^{c}} \Phi_{F} \left( \frac{L}{\xi^{F}_{var}(J_0) }  \right)
\label{fssferrowidth}
\end{eqnarray}
in terms of some correlation length
 \begin{eqnarray}
\xi^{F}_{var}(J_0)  \oppropto_{J_0 \to J_c^+} (J_0-J_c)^{-\nu^{F}_{var}}
\label{nuvarf}
\end{eqnarray}
yields the following behavior via the matching with Eq. \ref{wJferro}
 \begin{eqnarray}
\rho (J_0>J_c) && \propto
 \left[\xi^{F}_{var}(J_0)\right]^{\theta^c-\theta^{F}_{var}}
\label{fssferrowidthasymp}
\end{eqnarray}
i.e. the following relation between critical exponents
 \begin{eqnarray}
r=\nu^{F}_{var} (\theta^F_{var}-\theta^c)
\label{nufvar}
\end{eqnarray}

The previous values given for $\theta^c $, $ \theta^F_{var}$ and $r$ yields
 \begin{eqnarray}
\nu^{F}_{var}(d_{eff}=2) && \simeq \frac{0.28}{0.3-0.143} \simeq 1.8  \nonumber \\
\nu^{F}_{var}(d_{eff}=3) && \simeq \frac{0.39}{0.76-0.46 } \simeq 1.3
 \nonumber \\
\nu^{F}_{var}(d_{eff}=4) && \simeq   \frac{0.43}{1.24-0.87}   \simeq 1.16
 \nonumber \\
\nu^{F}_{var}(d_{eff}=5) && \simeq   \frac{0.44}{1.73-1.32} \simeq 1.07
 \nonumber \\
\nu^{F}_{var}(d_{eff}=6) && \simeq   \frac{0.45}{2.23-1.79} \simeq 1.02
\label{nuFvardiamond}
\end{eqnarray}
in agreement with Eq. \ref{nusgvardiamond} and Eq. \ref{nuFavdiamond}.

\begin{figure}[htbp]
\includegraphics[height=6cm]{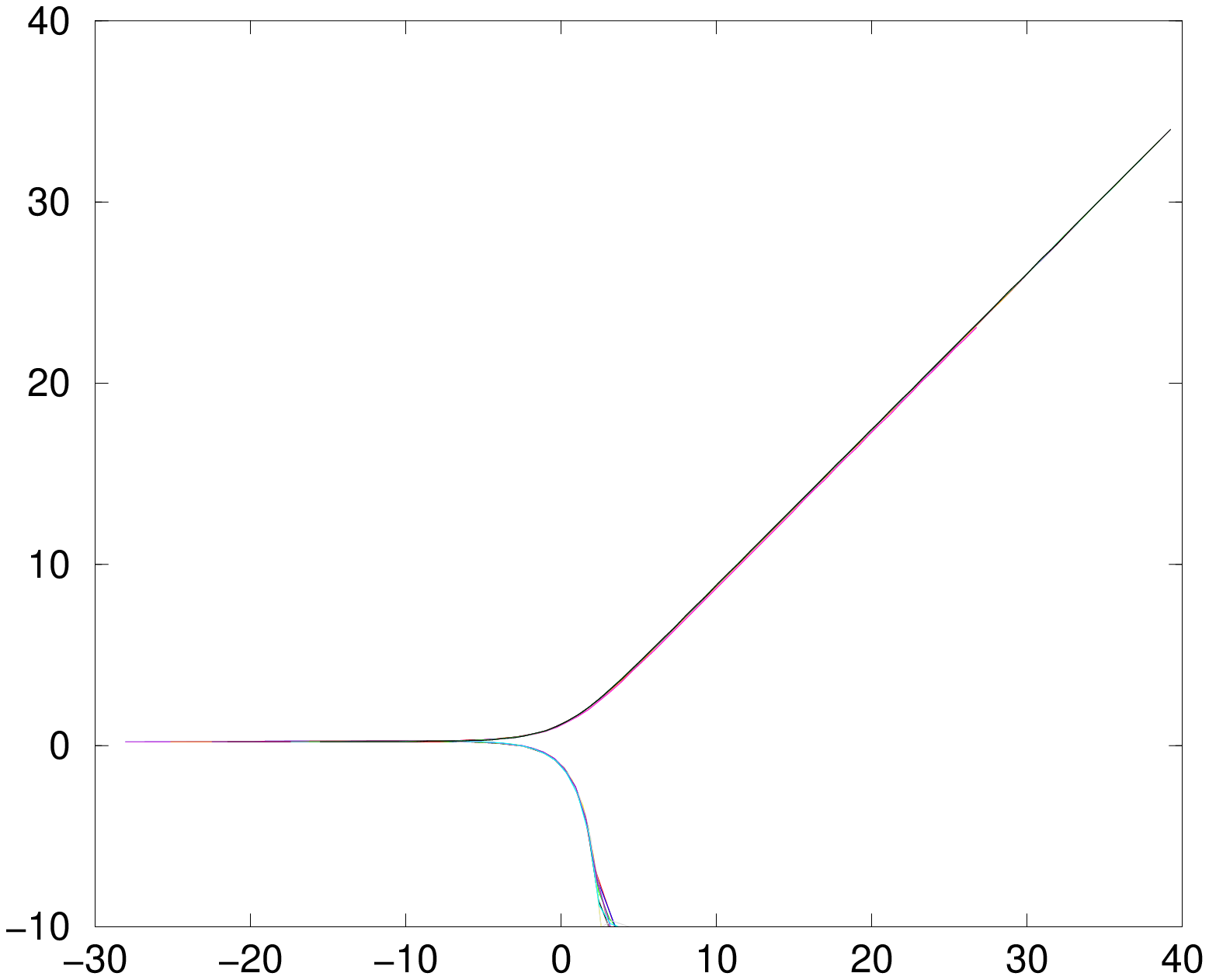}
\hspace{2cm}
\includegraphics[height=6cm]{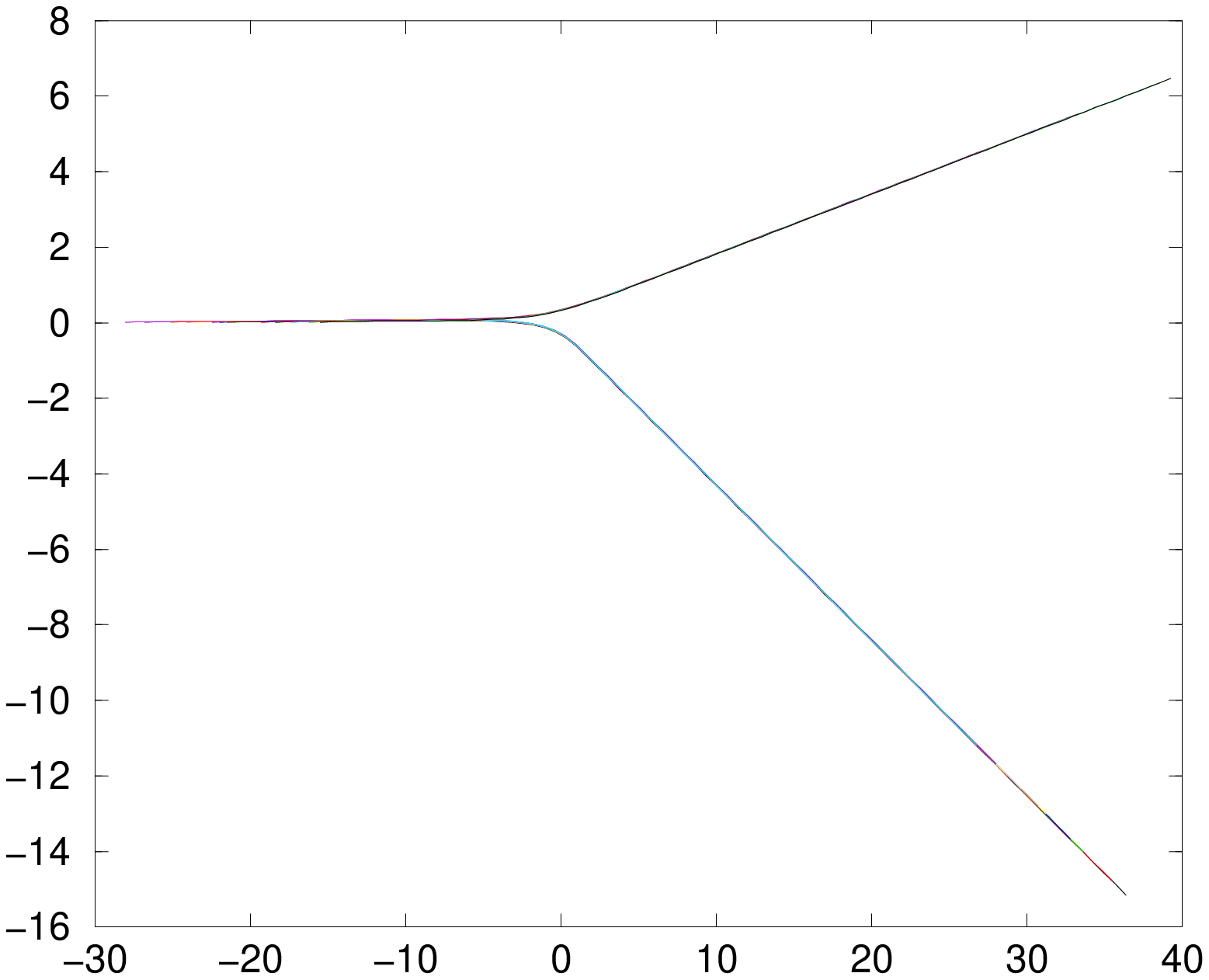}
\vspace{1cm}
\caption{ Data collapse of the RG flows for the case $d_{eff}=2$
(data of Figure \ref{figrgflow}) via finite-size scaling with the exponents
$\theta^c =0.14$ and $\nu=1.8$ : \\
(a)  $\ln \left( \frac{J_{av}^R(L)}{L^{\theta^c} }\right)$ as a function of
 $\ln \left( \vert J_0-J_c \vert L^{\frac{1}{\nu}} \right) $ (see Eq. \ref{fssferroav}) \\
(b)  $\ln \left( \frac{\Delta J^R (L)}{L^{\theta^c} }\right)$ as a function of
 $\ln \left( \vert J_0-J_c \vert L^{\frac{1}{\nu}} \right) $ (see Eqs \ref {fsssgwidth}
and \ref{fssferrowidth}).
}
\label{figfss}
\end{figure}

\subsection{ Conclusion  }

Our conclusion is thus that the average $J^R_{av}$ and the width $\Delta J^R $
of the distribution of renormalized coupling satisfy finite-size scaling
with a single correlation length exponent (Eqs \ref{nusgvardiamond},
\ref{nuFavdiamond} and \ref{nuFvardiamond})
\begin{eqnarray}
\nu=\nu^F_{av}=\nu^{F}_{var}=\nu^{SG}_{var}
\label{singlenu}
\end{eqnarray}
As an example, we show on Fig. \ref{figfss} 
the data collapse obtained from the data of 
Figure \ref{figrgflow}) concerning the RG flows for the case $d_{eff}=2$.

The role of the critical stiffness exponent $\theta^c$ 
in the relations between critical exponents can be summarized as follows
(Eqs \ref{nusgvar},  \ref{nufav} and \ref{nufvar})
 \begin{eqnarray}
y && = \nu (\theta^c-\theta^{SG}) \nonumber \\
s && =\nu (\theta^F_{av}-\theta^c) \nonumber \\
r && =\nu (\theta^F_{var}-\theta^c)
\label{fssrelations}
\end{eqnarray}
where $\theta^F_{av}=d_s=d_{eff}-1 $ in short-ranged models (Eq. \ref{avJferro}).

\section{Mean-Field Spherical Spin-Glass model }

\label{sec_sph}

\subsection{ Definition of the Model  }

In this section, we consider the fully connected Spherical Spin-Glass model introduced in \cite{sphericalSG} defined by the Hamiltonian 
\begin{eqnarray}
H=  - \frac{1}{2} \sum_{ i \ne j} J_{ij} S_i S_j
\label{Hsph}
\end{eqnarray}
where the $N$ spins are not 
Ising variables $S_i=\pm 1$ but are instead continuous variables $S_i \in ]-\infty,+\infty[$ submitted to the global constraint
\begin{eqnarray}
\sum_{i=1}^N S_i^2=N
\label{contraintesph}
\end{eqnarray}
Since each spin is connected to the other $(N-1)$ spins, the distribution of couplings of Eq. \ref{gauss} which was adapted to finite-connectivity lattices, \
has to be replaced by
the following rescaled random couplings
\begin{eqnarray}
 J_{ij} && = J_{ij}^{Ferro} +  J_{ij}^{(0)}  \nonumber \\
 J_{ij}^{Ferro} && = \frac{J_0}{N-1} \nonumber \\
 J_{ij}^{(0)} && = \frac{\epsilon_{ij}}{\sqrt{N-1}} 
\label{couplingsspherical}
\end{eqnarray}
where $\epsilon_{ij}= \epsilon_{ji}$ are drawn with the Gaussian distribution
of zero mean and unit variance
\begin{eqnarray}
P(\epsilon) =  \frac{1 }{\sqrt{ 2 \pi  } }  e^{ - \frac{ \epsilon^2 }{ 2  }}
\label{gausseps}
\end{eqnarray}
and where $J_0 \geq 0$ is the parameter controlling the ferromagnetic part of the coupling
\begin{eqnarray}
 \overline{J_{ij}}  =  J_{ij}^{Ferro} = \frac{J_0}{N-1} 
\label{j0sph}
\end{eqnarray}

\subsection{  Reminder on the ground state energy in each sample for $J_0=0$ } 

For $J_0=0$, the symmetric matrix ${\tilde J}^{(0)}$ of random couplings 
belongs to the Gaussian Orthogonal Ensemble (GOE).
Let us introduce its diagonalization
\begin{eqnarray}
{\tilde J}^{(0)} = \sum_{p=1}^N e_p \vert e_p >< e_p \vert
\label{diagoj}
\end{eqnarray}
in terms of the $N$ eigenvalues $e_p$ labeled in the order
\begin{eqnarray}
e_1>e_2>..>e_N
\label{diagoe}
\end{eqnarray}
and of the $N$ corresponding eigenvectors $\vert e_p > $.
It is convenient to write also the spin vector in this new basis
\begin{eqnarray}
\vert S > = \sum_{i=1}^N S_i \vert i > = \sum_{p=1}^N S_{e_p} \vert e_p >
\label{sdiagoj}
\end{eqnarray}
The energy of Eq. \ref{Hsph} and the spherical constraint of Eq. \ref{contraintesph}
then read
\begin{eqnarray}
H && = - \frac{1}{2} \sum_{p=1}^{N} e_p S_{e_p}^2 
\nonumber \\
   N && = \sum_{p=1}^N S_{e_p}^2 
\label{zsphdiag}
\end{eqnarray}
The ground-state is then obvious : the minimal energy is obtained by putting
 the maximal possible weight in the maximal eigenvalue $e_1$
(Eq. \ref{diagoe})
and zero weight in all other eigenvalues $e_p$ with $p=2,3,..,N$
\begin{eqnarray}
S_{e_p \ne e_1}^{GS} && = 0
\nonumber \\
(S_{e_1}^{GS})^2 && = N
\label{eigengs}
\end{eqnarray}
The corresponding ground-state energy 
\begin{eqnarray}
E^{GS}(N)  = - N \frac{ e_1 }{2}
\label{egssph}
\end{eqnarray}
thus only involves the maximal eigenvalue $e_1$ of the GOE matrix.
Its asymptotic distribution for large $N$ is known to be 
\begin{eqnarray}
 e_1 = 2  \left( 1- \frac{u}{2 N^{2/3}} \right)
\label{lambdamaxTW}
\end{eqnarray}
where the value $2 $ corresponds to the boundary of the semi-circle law 
\begin{eqnarray}
\rho(e) =  \frac{1}{2 \pi} \sqrt {4-e^2}
\label{semicircle}
\end{eqnarray}
that emerges in the thermodynamic limit
$N \to +\infty$, and where $u$ is a random variable of order $O(1)$
distributed with the Tracy-Widom distribution \cite{tracyWidom}.
The ground-state energy of Eq. \ref{egssph} thus reads \cite{andreanov,us_sgoverlaptyp}
\begin{eqnarray}
E^{GS}(N)= - N \frac{ e_1 }{2} = -  N + N^{\frac{1}{3}} \frac{ u}{2 }
\label{egssphfinal}
\end{eqnarray}
In summary for $J_0=0$, the extensive term is non-random, and the next subleading term
is of order $N^{\frac{1}{3}}$ and random, distributed with
 the Tracy-Widom distribution.
So the droplet exponent $\omega^{SG } $ characterizing the spin-glass phase
takes the simple value
\begin{eqnarray}
\omega^{SG }_{sph} && =\frac{1}{3}
\label{thetasph}
\end{eqnarray}
when redefined in terms of the number $N$ of spins (and not with respect to the length
which does not exist in such fully connected models).

\subsection{  Analysis for $J_0>0$ } 

\subsubsection{ Reformulation as a localized impurity effect}

As explained in Ref. \cite{sphericalSG}, the case $J_0>0$ can be analyzed
by diagonalizing the ferromagnetic matrix (Eq. \ref{couplingsspherical})
\begin{eqnarray}
 J_{ij}^{Ferro}  = \frac{J_0}{N-1} (1-\delta_{i,j})
\label{jijferro}
\end{eqnarray}
The ``ferromagnetic eigenvector''
\begin{eqnarray}
\vert a_0 > \equiv \frac{1}{\sqrt N} \sum_{i=1}^N \vert i >
\label{v0ferro}
\end{eqnarray}
 is the eigenstate of the matrix of Eq. \ref{jijferro} with 
the maximal eigenvalue $J_0$
\begin{eqnarray}
 {\tilde J}^{Ferro} \vert a_0 > = J_0 \vert a_0 > 
\label{ferroeigen}
\end{eqnarray}
and the matrix of Eq. \ref{jijferro} can be then
decomposed using the projector onto the ferromagnetic eigenvector
and the projector onto the orthogonal space
\begin{eqnarray}
 J^{Ferro} = J_0 \vert a_0 > <a_0 \vert
 - \frac{J_0}{N-1} \left(1 - \vert a_0 > <a_0 \vert \right)
\label{ferromartix}
\end{eqnarray}
So the orthogonal space to the ``ferromagnetic eigenvector''
of Eq. \ref{v0ferro} is associated to the degenerate eigenvalue
 $(-\frac{J_0}{N-1})$ that can be neglected in the following  \cite{sphericalSG}.
Using some basis $\vert a_k >$ with $k=1,2,..,N-1$
in the orthogonal space to $\vert a_0 > $, 
the random couplings of Eq. \ref{couplingsspherical}
become
\begin{eqnarray}
J_{k,q} && = J_{k,q}^{Ferro}+J_{k,q}^{(0)} \nonumber \\
 J_{k,q}^{Ferro} && = J_0 \delta_{k=0,q=0} \nonumber \\
J_{k,q}^{(0)} && = \frac{\epsilon_{k,q}}{\sqrt{N-1}}
\label{jeffectif}
\end{eqnarray}
where the $\epsilon_{k,q}$ are Gaussian random variables of zero mean and variance unity as in Eq. \ref{gausseps}. So $J^{(0)}$ is a GOE random matrix with its associated Green
function
\begin{eqnarray}
G_0(z) \equiv \frac{1}{z-J^{(0)}} = \sum_{p=1}^N \frac{\vert e_p> < e_p \vert }{z-e_p}
\label{greenzero}
\end{eqnarray}
where $e_p$ are the energy levels of the GOE matrix (Eq. \ref{diagoj})
and $\vert e_p> $ the corresponding eigenvectors.
Then the full matrix $ J_{k,q}$ of Eq. \ref{jeffectif} corresponds
to the problem where a single impurity localized on the vector $\vert a_{k=0} >$ 
is added to the GOE matrix $J_{k,q}^{(0)} $.
This problem is exactly soluble as follows \cite{izyumov} : 
the Green function in the presence of the impurity
\begin{eqnarray}
G(z) \equiv \frac{1}{z-J}= \sum_{n=1}^N \frac{\vert E_n> < E_n \vert }{z-E_n}
\label{green}
\end{eqnarray}
where $E_n$ are the energy levels in the presence of the impurity,
and $\vert E_n> $ the corresponding eigenvectors, satisfies the Dyson equation
as an operator identity
 \begin{eqnarray}
G(z) =G_0(z) +  G_0(z) J^{Ferro} G(z)
\label{dyson}
\end{eqnarray}
with the formal solution
 \begin{eqnarray}
G(z) = \frac{1}{1-G_0(z) J^{Ferro} } G_0(z) =G_0(z)+  G_0(z) J^{Ferro} G_0(z) + G_0(z) J^{Ferro} G_0(z)  J^{Ferro} G_0(z)
+...
\label{dysonsol}
\end{eqnarray}
The simplification comes from the form of the perturbation operator
\begin{eqnarray}
J^{Ferro}=J_0  \vert a_0 > <a_0 \vert
\label{perturbation}
\end{eqnarray}
 that allows to resum the infinite series of Eq. \ref{dysonsol}
into
 \begin{eqnarray}
G(z) = G_0(z)+ J_0 \frac{ G_0(z)\vert a_0 >
 <a_0 \vert G_0(z)}{1 -J_0 <a_0 \vert G_0(z)  \vert a_0 > }
\label{locsimpli}
\end{eqnarray}
It is thus convenient to introduce the local unperturbed Green function on the ferromagnetic eigenvector $\vert a_0 >$ (Eq \ref{greenzero}) 
 \begin{eqnarray}
g_0(z) \equiv  <a_0 \vert G_0(z)  \vert a_0 > 
= \sum_{p=1}^N \frac{ \vert  <a_0\vert e_p> \vert^2 }{z-e_p}
\label{localg0}
\end{eqnarray}
Since the ferromagnetic eigenvector $\vert a_0 >$ is an arbitrary vector with respect to the unperturbed GOE matrix $J^{(0)}$ that has only delocalized eigenvectors $\vert e_p>$, one has $\vert  <a_0\vert e_p> \vert^2=1/N$ so that it can be approximated by
 \begin{eqnarray}
g_0(z) \simeq \frac{1}{N} \sum_{p=1}^N \frac{ 1 }{z-e_p}
\label{localg0approx}
\end{eqnarray}
that contains only the GOE energies $e_p$.

The local perturbed Green function on the ferromagnetic eigenvector $\vert a_0 >$ 
 \begin{eqnarray}
g(z) \equiv  <a_0 \vert G(z)  \vert a_0 > 
= \sum_{n=1}^N \frac{  \vert  <a_0 \vert E_n> \vert^2 }{z-E_n}
\label{localg}
\end{eqnarray}
can be now directly computed from $g_0(z)$ of Eq. \ref{localg0approx}
using Eq. \ref{locsimpli},
\begin{eqnarray}
g(z) = g_0(z)+  J_0 \frac{ g_0^2(z) }{1 -J_0 g_0(z) }
\label{localgeq}
\end{eqnarray}

\subsection{ Results in the thermodynamic limit $N \to +\infty$ }

In the thermodynamic limit $N \to +\infty$, 
Eq. \ref{localg0approx} can be explicitly computed from
the semi-circle law of Eq. \ref{semicircle}
 \begin{eqnarray}
g_0^{(N=+\infty)}(z) = \frac{1}{2 \pi}
 \int_{-2}^{+2} de \frac{\sqrt {4-e^2}}{z-e} = \frac{ z - \sqrt{z^2-4}}{2 }
\label{localg0semicircle}
\end{eqnarray}
Then Eq. \ref{localgeq} will have a pole at an energy $z=E_F>2$ above the semi-circle
law if 
 \begin{eqnarray}
0= 1- J_0 g_0^{(N=+\infty)}(E_F) = 1-J_0 \frac{ E_F - \sqrt{E_F^2-4}}{2 }
\label{localg0semicircleeq}
\end{eqnarray}
leading to
 \begin{eqnarray}
\sqrt{E_F^2-4}  =   E_F -\frac{2}{J_0} 
\label{eqef}
\end{eqnarray}
For $J_0<1$, there is no solution, whereas for $J_0>1$, 
 the following solution exists \cite{sphericalSG}
 \begin{eqnarray}
E_F(J_0>1)= J_0+\frac{1}{J_0}
\label{soluthermo}
\end{eqnarray}
and the corresponding residue
 reads using the explicit form of Eq. \ref{localg0semicircle}
\begin{eqnarray}
\vert <E_0 \vert a_0>\vert^2 && = J_0 \frac{ g_0^2(E_F) }{ (-J_0 g_0'(E_F)) }
= 1- \frac{1}{J_0^2}
\label{localgeqresidue}
\end{eqnarray}

In summary, the intensive energy of the ground state remains frozen to Eq. \ref{egssph}
for $J_0<1$
\begin{eqnarray}
\frac{E^{GS}_N(J_0 < 1)}{N} \operarrow_{N \to +\infty}  - 1
\label{egssphsg}
\end{eqnarray}
whereas for $J_0>1$, the intensive energy of the ground state is governed by the ferromagnetic pole of Eq. \ref{soluthermo}
\begin{eqnarray}
\frac{E^{GS}_N(J_0 > 1)}{N} \operarrow_{N \to +\infty}  - \frac{  J_0+\frac{1}{J_0} }{2}
= -1 - \frac{(J_0-1)^2}{2 J_0^2} 
\label{egssphsgferro}
\end{eqnarray}
So the singularity in $(J_0-J_c)^{2-\alpha}$ involves the standard mean-field exponent
\cite{sphericalSG}
\begin{eqnarray}
\alpha_{sph}=0
\label{alphasph}
\end{eqnarray}
The corresponding intensive magnetization (Eq. \ref{localgeqresidue})
\begin{eqnarray}
m_N(J_0>1) \operarrow_{N \to +\infty} \vert <E_0 \vert a_0>\vert =  \frac{\sqrt{J_0^2-1} }{J_0}
\label{magnsph}
\end{eqnarray}
involves the singularity $(J_0-J_c)^{\beta}$ with the standard mean-field exponent
\cite{sphericalSG}
\begin{eqnarray}
\beta_{sph}=\frac{1}{2}
\label{betasph}
\end{eqnarray}
In summary, in the thermodynamic limit $N \to +\infty$, 
the singularities of intensive observables
are governed by the standard mean-field exponents  \cite{sphericalSG}.
But it is interesting to discuss now the finite-size effects.

\subsection{ Finite-size effects in the two phases and at criticality }

(i) In the spin-glass phase,  we have already seen in Eq. \ref{egssphfinal}
that the subleading random term with respect to the thermodynamic limit of Eq. \ref{egssphsg} is governed by the exponent $\omega^{SG}_{sph}=1/3$
\begin{eqnarray}
E^{GS}_N(J_0<1) &&  = -  N + N^{\omega^{SG}_{sph}} \frac{ u}{2 } \nonumber \\
\omega^{SG}_{sph}&& =\frac{1}{3}
\label{egssphfinaln}
\end{eqnarray}

(ii) In the random ferromagnetic phase, $G_0(z)$
 of Eq. \ref{localg0approx}
displays fluctuations of order $1/N^{1/2}$ with respect to its thermodynamic limit
of Eq. \ref{localg0semicircle}, so we expect that the pole $E_F$ of Eq. \ref{soluthermo} will inherit from these fluctuations. So for the ground state energy,
the leading finite-size correction with respect to the thermodynamic limit of Eq. \ref{egssphsgferro} will be
\begin{eqnarray}
E^{GS}_N(J_0 > 1) && \simeq  - N \frac{  J_0+\frac{1}{J_0} }{2}+N^{\frac{1}{2}} v
\nonumber \\
\omega^{F}_{sph}&& =\frac{1}{2}
\label{egssphsgferron}
\end{eqnarray}

(iii) Near the critical point for finite $N$, we need to discuss 
the equation for the pole $E_F$ when the GOE energy levels $e_p$ are still discrete,
 \begin{eqnarray}
\frac{1}{J_0} =   g_0^{(N)}(E_F) =\frac{1}{N} \sum_{p=1}^N \frac{ 1 }{E_F-e_p}
\label{eqresodiscrete}
\end{eqnarray}
When the pole $E_F$ is very close to the highest GOE energy $e_1$,
the corresponding term will dominate the sum in Eq. \ref{eqresodiscrete}
to yield the solution
 \begin{eqnarray}
E_F \simeq e_1+\frac{J_0}{N}
\label{solresodiscrete}
\end{eqnarray}
Using the Tracy-Widom scaling for the gap $(e_1-e_2) \propto N^{-2/3}$,
one obtains that the biggest 
subleading term in Eq. \ref{eqresodiscrete} is then of order
 \begin{eqnarray}
\frac{1}{N (E_F-e_2)}=\frac{1}{N (E_F-e_1+e_1-e_2)}=\frac{1}{J_0+u N^{1/3}} \propto N^{-1/3}
\label{eqresodiscretenext}
\end{eqnarray}
and can be indeed neglected.
So we expect that at criticality, the finite-size behavior of the ground state energy 
\begin{eqnarray}
E^{GS}_N(J_0=1) \simeq  = -  N + N^{\omega^c_{sph}} w
\label{egssphcritin}
\end{eqnarray}
is governed by 
the same exponent 
\begin{eqnarray}
\omega^c_{sph} =\frac{1}{3}
\label{thetacsph}
\end{eqnarray}
as in the SG phase of Eq. \ref{egssphfinaln}.

\subsection{ Scaling of the renormalized coupling $J^R_N$ }

In fully connected models of $N$ spins, the notion of renormalized coupling of
 Eq. \ref{defjr} can be adapted as
\begin{eqnarray}
J^R_N \equiv  E_{GS}^{(AP)}(N)-E_{GS}^{(P)}(N) 
\label{defjrmf}
\end{eqnarray}
where 'Periodic' and 'Antiperiodic' are defined by
 the following prescription introduced for long-ranged spin-glasses
on a circle \cite{KY} :
for each disordered sample $(J_{ij})$ considered as 'Periodic', the 'Antiperiodic'
consists in changing the sign $J_{ij} \to -J_{ij}$ for all pairs $(i,j)$ where
the shortest path on the circle goes through the bond $(L,1)$.

For the spherical spin-glass model at $J_0=0$, and more generally in the whole spin-glass phase $J_0<J_c=1$, the width $\Delta J^R_N $ of the probability distribution
of renormalized coupling $J^R(N) $ grows as \cite{KY,us_sgoverlaptyp}
\begin{eqnarray}
\Delta J^R (J_0<J_c) && \simeq \Upsilon(J_0) N^{\omega^{SG}_{sph}}  \nonumber \\
\omega^{SG}_{sph} && = \frac{1}{3}
\label{sgjrmf}
\end{eqnarray}
because the leading non-random extensive term of Eq. \ref{egssphfinaln}
cancels in the difference between Periodic and Antiperiodic in Eq. \ref{defjrmf},
so that the subleading random term of Eq. \ref{egssphfinaln} becomes the leading term in Eq. \ref{sgjrmf}.

In the ferromagnetic phase $J_0>J_c=1$, 
the averaged renormalized coupling $J^R_{av} (J_0>J_c) $ presents the same scaling as the pure ferromagnet : for short-ranged models, the energy cost of an interface grows as the surface $L^{d_s}$ of a system-size interface (Eqs \ref{avJferro} and \ref{defdsSR}).
But in fully connected model, the surface dimension $d_s$ becomes equal to the volume dimension $d$, so that the energy cost scales with the total number $N=L^d$ of spins
 \begin{eqnarray}
 J^R_{av} (J_0>J_c) \propto \sigma(J_0) N
\label{avJferrosph}
\end{eqnarray}
i.e. here the leading extensive term of Eq. \ref{egssphsgferron} does not vanish
between Periodic and Antiperiodic which contains antiferromagnetic components.
As a consequence, the singularity of $\sigma(J_0)$ of Eq. \ref{critisigma}
is expected to involve the same exponent as the exponent $(2-\alpha_{sph})=2$
of the intensive energy
 \begin{eqnarray}
 s_{sph}=2
\label{s2sph}
\end{eqnarray}

The width of the distribution around this average is expected to scale
with the fluctuation exponent of Eq. \ref{egssphsgferron}
\begin{eqnarray}
\Delta J^R (J_0>J_c) && \simeq \rho(J_0)  N^{\omega^{F}_{sph}}  \nonumber \\
\omega^{F}_{sph} && = \frac{1}{2}
\label{wJferrojrmf}
\end{eqnarray}

At criticality, both the averaged value and the width display the same scaling 
 \begin{eqnarray}
 J^R_{av} (J_0=J_c) \propto  N^{\omega^c_{sph}} \propto \Delta J^R (J_0=J_c)
\label{jcritisph}
\end{eqnarray}
with the critical exponent $\omega^c_{sph} =\frac{1}{3}$ of Eq. \ref{thetacsph}.

The fact that $\omega^c_{sph} =\frac{1}{3}$ and $\omega^{SG}_{sph}=\frac{1}{3}$
coincide implies that the 
stiffness modulus $\Upsilon(J_0)$ of Eq. \ref{sgjrmf} is not singular at the transition,
so that the critical exponent of Eq. \ref{dvupsilon} vanishes
 \begin{eqnarray}
y_{sph}=0
\label{ymf}
\end{eqnarray}
This behavior seems to represent well what happens in finite dimension 
for large enough $d$ (Eq. \ref{dvupssgdiamond}). 

On the ferromagnetic side, the finite-size scaling form
with respect to the number $N$ of spins
for the averaged coupling
 \begin{eqnarray}
J^R_{av} (J_0>J_c) =  N^{\omega^c_{sph}} \Phi_{av} \left[(J_0-J_c) N^{\frac{1}{\mu}} \right]
\label{fssjavsph}
\end{eqnarray}
has to match Eq. \ref{avJferrosph} with Eq. \ref{s2sph}
so that 
 \begin{eqnarray}
\mu=3
\label{mumfav}
\end{eqnarray}
in agreement with the finite-size scaling exponent found
for the ferromagnetic/spin-glass transition on the Bethe lattice \cite{liers}.
Then the finite-size scaling form
for the width of the distribution of the average coupling 
 \begin{eqnarray}
\Delta J^R (J_0>J_c) =  N^{\omega^c_{sph}} \Phi_{var} \left[(J_0-J_c) N^{\frac{1}{\mu}} \right]
\label{fssjwsph}
\end{eqnarray}
has to match Eq. \ref{wJferrojrmf}, so that the critical exponent of Eq. \ref{critirho}
reads
 \begin{eqnarray}
r_{sph}=\frac{1}{2}
\label{rmf}
\end{eqnarray}
This value seems to describe well  what happens in finite dimension 
for large enough $d$ (Eq. \ref{critirhodiamond}).

\section{ Conclusion}

\label{sec_conclusion}

For the Ising model with Gaussian random coupling of average $J_0$ and unit variance,
we have characterized the zero-temperature spinglass-ferromagnetic transition as a function of the control parameter $J_0$ via the size-$L$ dependent renormalized coupling defined as the domain-wall energy $J^R(L) \equiv  E_{GS}^{(AF)}(L)-E_{GS}^{(F)}(L)$.
In the first part of the paper, we have studied numerically the critical exponents for the average and the width of the probability distribution of this renormalized coupling within the Migdal-Kadanoff approximation as a function of the dimension $d=2,3,4,5,6$. 
In the second part of the paper, we have compared with
 the corresponding mean-field exponents for spherical model.
Our main conclusions are the following :

(i) the critical stiffness exponent $\theta^c>0$ is the main signature of the zero-temperature nature of the transition (whereas thermal transitions towards the paramagnetic phase correspond to $\theta^c=0$) and appear in the finite size scaling relations between exponent as summarized in Eq. \ref{fssrelations}.

(ii) in low dimensions, the critical stiffness exponent $\theta^c$ is clearly bigger than the spin-glass stiffness exponent $\theta^{SG}$, but they turn out to coincide in high enough dimension and in the mean-field spherical model.

We hope that in the future, the critical stiffness exponent $\theta^c$
will be measured for hypercubic lattice in dimensions $d>2$ (see 
\cite{rieger,amoruso_hartmann} for measures in $d=2$), and for the 
one-dimensional long ranged spin-glass in order to compare with the values of $\theta^{SG}$ measured in \cite{KY}.

\appendix

\section{ Reminder on the Random Energy Model }

\label{app_rem}

The Random Energy Model is a mean-field spin-glass model that has been introduced and solved in \cite{derrida_rem}. In this Appendix, we recall only its properties at zero-temperature (see \cite{derrida_rem} for full calculations at non-zero temperature).

\subsection{ Properties of the spin-glass phase for $J_0=0$ }

A realization of the Random Energy Model of $N$ spins 
is defined by the set of $2^N$ independent random
energies levels $E_i$ drawn with the Gaussian distribution
 \begin{eqnarray}
G_N(E) = \frac{1}{\sqrt{ \pi N }} e^{- \frac{E^2}{ N } }
\label{gaussrem}
\end{eqnarray}
The ground-state energy $E^{GS}_N$ is simply
the minimal energy of these $2^N$ independent levels
 \begin{eqnarray}
E^{GS}_N = {\rm min} (E_1,E_2,.,E_{2^N})
\label{mingaussrem}
\end{eqnarray}
This standard problem of extreme value statistics \cite{Gum_Gal} 
can be solved by considering the following integral of
its probability distribution ${\cal P}(E^{GS}_N)$ 
 \begin{eqnarray}
 \int_x^{+\infty} dE^{GS}_N {\cal P}(E^{GS}_N)
= Prob(x \leq E^{GS}_N)   && =\prod_{i=1}^{2^N}
Prob(x \leq E_i)
= \left[ \int_{x}^{+\infty} dE G_N(E)  \right]^{2^N}
\nonumber \\
&& \opsimeq_{N \to +\infty}  e^{ 2^N \ln \left[ 1-\int_{-\infty}^x  dE G_N(E) \right]  }
 \simeq e^{ - 2^N \int_{-\infty}^x  dE G_N(E)   }
\label{pegsrem}
\end{eqnarray}
Using the complementary error function and its asymptotic expansion at infinity
 \begin{eqnarray}
{\rm erfc}(z) \equiv \frac{2}{\sqrt{\pi}} \int_{z}^{+\infty} dt e^{-t^2}
\opsimeq_{z \to +\infty} \frac{1}{z \sqrt{\pi} } e^{-z^2}
\label{erfc}
\end{eqnarray}
one obtains in the regime of interest $x \to -\infty$
 \begin{eqnarray}
\int_{-\infty}^x  dE G_N(E) && = \int_{-\infty}^x \frac{dE }{\sqrt{ \pi N }}
 e^{- \frac{E^2}{N } } = 
\int_{-\infty}^{\frac{x}{\sqrt{N}}} \frac{dy }{\sqrt{ \pi}} e^{- y^2}
=\frac{1}{2} {\rm erfc} \left(- \frac{x}{\sqrt{N}} \right)
 \simeq \frac{1}{\left(- \frac{x}{\sqrt{N}} \right)  2 \sqrt{\pi} } e^{- \frac{x^2}{N } }
\label{errorfunction}
\end{eqnarray}
yielding
 \begin{eqnarray}
 \int_x^{+\infty} dE^{GS}_N {\cal P}(E^{GS}_N)
 \opsimeq
 \simeq e^{ - 2^N  \frac{\sqrt{N} e^{- \frac{x^2}{N } }  }{(- x)  2 \sqrt{\pi} } }
\label{pegsremfin}
\end{eqnarray}
So the characteristic scale $x_N$ where the argument of the exponential is unity
reads asymptotically for large $N$
 \begin{eqnarray}
 x_N  =  - N \sqrt{\ln 2} + \frac{ \ln N  }{ 4  \sqrt{\ln 2}}
+ \frac{ \ln \left(    2  \sqrt{\pi } \ln 2   \right) }{ 2  \sqrt{\ln 2}}
 +o(1)
\label{xetoile}
\end{eqnarray}
Making the change of variable
 \begin{eqnarray}
E^{GS}_N= x_N +v
\label{Exetoileu}
\end{eqnarray}
one obtains that the variable $v$ is a $O(1)$ random variable
  distributed asymptotically with the Gumbel distribution $g(v)$ \cite{Gum_Gal} 
 \begin{eqnarray}
\int_u^{+\infty} dv g(v) \simeq e^{- e^{2 \sqrt{\ln 2} u}}
\label{gumbel}
\end{eqnarray}
So in the Random Energy Model, the spin-glass phase is characterized by 
a logarithmic correction to extensivity for the ground state energy (Eq. \ref{xetoile}), i.e. the droplet exponent vanishes
 \begin{eqnarray}
\omega^{SG}_{REM}=0
\label{thetarem}
\end{eqnarray}
in contrast to Eq. \ref{thetasph} concerning the spherical mean-field model.

\subsection{ Properties in the presence of an averaged coupling $J_0>0$ }

In the presence of some averaged ferromagnetic coupling $J_0$,
the Random Energy Model is generalized as follows (see section VIII of Ref. \cite{derrida_rem}) :

(i) among the $2^N$ independent 
energy levels, $\binom{N}{\frac{N+M}{2}}$ have magnetization $M$,
where  $M=-N,..,+N$. 

(ii) the energy of a level with magnetization $M$ is drawn
 with the magnetization-dependent distribution
 \begin{eqnarray}
G_{M,N}(E) = \frac{1}{\sqrt{ \pi N }} e^{- \frac{1}{N } \left(E+J_0 \frac{M^2}{2N}  \right)^2}
\label{pme}
\end{eqnarray}

So Eq. \ref{pegsrem} for the probability distribution
of the ground state energy  ${\cal P}(E^{GS}_N)$  becomes
 \begin{eqnarray}
\int_x^{+\infty} dE^{GS}_N {\cal P}(E^{GS}_N) && =Prob(x \leq E^{GS}_N) 
= \prod_{M=-N}^{+N} \left[ \int_{x}^{+\infty} dE G_{M,N}(E)  \right]^{\binom{N}{\frac{N+M}{2}}}
\nonumber \\
&& \simeq e^{ \sum_{M=-N}^{+N} \binom{N}{\frac{N+M}{2}} \ln \left[ 1-\int_{-\infty}^x  dE G_{M,N}(E) \right]  }
 \simeq e^{ - \sum_{M=-N}^{+N} \binom{N}{\frac{N+M}{2}} \int_{-\infty}^x  dE G_{M,N}(E)   }
\label{pegsminav}
\end{eqnarray}
Using the complementary error function of Eq \ref{erfc}
and the Stirling approximation for the binomial coefficient
 \begin{eqnarray}
\binom{N}{\frac{N+Nm}{2}} = \frac{N!}{\left(N \frac{1+m}{2}\right)! 
\left( N \frac{1-m}{2}\right)!}= 2^N \frac{1}{\sqrt{2 \pi N}}
 e^{- \frac{N}{4} \left[ (1+m)\ln (1+m)+  (1-m)\ln (1-m)\right]}
\label{binomial}
\end{eqnarray}
one obtains that the characteristic scale $x_{N,M=Nm}$ 
where the argument of the exponential in Eq. \ref{pegsminav}
is unity, reads asymptotically for large $N$ at fixed intensive magnetization $m$
 \begin{eqnarray}
x^*_{N,M=Nm} \simeq - N \sqrt{\ln 2-\phi(m)}- N J_0  \frac{m^2}{2 }
 + \frac{\ln N}{2 \sqrt{\ln 2-\phi(m)} }+O(1)
\label{xetoilem}
\end{eqnarray}
with
 \begin{eqnarray}
\phi(m)= \frac{1}{2} \left[ (1+m)\ln (1+m)+  (1-m)\ln (1-m)\right] \opsimeq_{m \to 0}
\frac{m^2}{2}+\frac{m^4}{12}+O(m^6)
\label{phim}
\end{eqnarray}

So the intensive energy $e(m)=x^*_{N,M=Nm}/N$ as a function of the intensive magnetization $m$ reads
 \begin{eqnarray}
e(m) \equiv -  \sqrt{\ln 2-\phi(m)}-  J_0  \frac{m^2}{2 }
\label{efunctionm}
\end{eqnarray}
with the following expansion near zero magnetization
 \begin{eqnarray}
e(m) = -  \sqrt{\ln 2} + \left(\frac{1}{2 \sqrt{\ln 2}} -  J_0 \right) \frac{m^2}{2 } + \frac{ 11 }{192 \sqrt{\ln 2} } m^4 +O(m^6)
\label{efunctionmzero}
\end{eqnarray}
The critical value $J_c$ correspond to the value of $J_0$ where the coefficient
of the quadratic term of the magnetization changes sign  \cite{derrida_rem}
 \begin{eqnarray}
J_c= \frac{1}{2 \sqrt{\ln 2}}
\label{jccritirem}
\end{eqnarray}
For $J_0<J_c$, the function $e(m)$ is minimum at $m=0$,
so the ground-state has for intensive parameters \cite{derrida_rem}
 \begin{eqnarray}
m_{GS}(J_0<J_c) && =0 \nonumber \\
e_{GS}(J_0<J_c) && = -  \sqrt{\ln 2}
\label{gsrembelow}
\end{eqnarray}

For $J_0>J_c$, the minimum of the function $e(m)$ is not at $m=0$ anymore,
but at two symmetric values $\pm m_{GS}$ where
 \begin{eqnarray}
0=e'(m_{GS}) = \left( J_c -  J_0 \right) m + \frac{ 11 }{48 \sqrt{\ln 2} } m^3 +O(m^5)
\label{deriefunctionm}
\end{eqnarray}
No near the transition one obtains the the standard thermodynamic mean-field
exponents
 \begin{eqnarray}
m_{GS}(J_0>J_c) && \propto \left( J_c -  J_0 \right)^{\beta_{REM}} \ \ {\rm with } \ \beta_{REM}=\frac{1}{2} \nonumber \\
e_{GS}(J_0>J_c)+ \sqrt{\ln 2}  && \propto  \left( J_c -  J_0 \right)^{2-\alpha_{REM}} \ \ {\rm with } \ \alpha_{REM}=0
\label{gsremabove}
\end{eqnarray}
as in the mean-field spherical model (Eqs \ref{alphasph} and \ref{betasph}).
Exactly at criticality, the leading finite-size correction for the ground state energy is again logarithmic (Eq. \ref{xetoilem}), so the 
 the critical droplet exponent vanishes
 \begin{eqnarray}
\omega^{c}_{REM}=0
\label{thetacrem}
\end{eqnarray}
 and coincides with
the droplet exponent of the spin-glass phase $\omega^{SG}_{REM}=0 $
 (Eq. \ref{thetarem}).
Finally, the finite-size scaling near criticality $(J_0-J_c)N^{\mu}$
is governed by the exponent
 \begin{eqnarray}
\mu_{REM}=2
\label{murem}
\end{eqnarray}
which differs from the value of Eq. \ref{mumfav} found for the spherical mean-field model.

\end{document}